\documentclass[10pt, conference]{IEEEtran}

\ifCLASSINFOpdf
\else
\fi

\usepackage{amssymb}
\usepackage{amsmath}
\usepackage[version=3]{mhchem}
\usepackage{graphicx}
\usepackage{multirow}
\usepackage{wrapfig}
\usepackage{color}
\usepackage{colortbl}
\usepackage{soul}
\usepackage{subfigure}
\usepackage{enumerate}
\usepackage{paralist}
\usepackage[normalem]{ulem}

\begin{document}
\title{Modeling and Analysis of SiNW BioFET as \\Molecular Antenna for Bio-Cyber Interfaces towards\\ the Internet of Bio-NanoThings}
\author{
\IEEEauthorblockN{\textbf{Murat Kuscu~~~~~~~Ozgur B. Akan}}
\IEEEauthorblockA{Next-generation and Wireless Communications Laboratory (NWCL)\\
Department of Electrical and Electronics Engineering\\
Koc University, Istanbul, Turkey \\ Email: \{mkuscu, akan\}@ku.edu.tr}
}

\maketitle

\begin{abstract}
Seamless connection of molecular nanonetworks to macroscale cyber networks is envisioned to enable the \emph{Internet of Bio-NanoThings}, which promises for cutting-edge applications, especially in the medical domain. The connection requires the development of an interface between the biochemical domain of molecular nanonetworks and the electrical domain of conventional electromagnetic networks. To this aim, in this paper, we propose to exploit field effect transistor based biosensors (bioFETs) to devise a molecular antenna capable of transducing molecular messages into electrical signals. In particular, focusing on the use of SiNW FET-based biosensors as molecular antennas, we develop deterministic and noise models for the antenna operation to provide a theoretical framework for the optimization of the device from communication perspective. We numerically evaluate the performance of the antenna in terms of the Signal-to-Noise Ratio (SNR) at the electrical output.
\end{abstract}


\IEEEpeerreviewmaketitle

\section{Introduction}
\emph{Internet of Bio-NanoThings (IoBNT)} describes the vision of connecting the networks of biological-nanoscale functional entities, e.g., bacterial colonies, synthetic and natural cells, artificial implants, with each other and with cyber-networks, e.g., the Internet \cite{Akyildiz2015} \cite{Kuscu2015}. The IoBNT is expected to significantly extend the coverage of Internet of Things (IoT) and enable promising applications such as continuous health monitoring and bacterial sensor-actor networks inside human body.

Biological entities naturally communicate with each other by using molecules to encode, transmit and receive information. This \emph{Molecular Communication (MC)} paradigm has been extensively studied in the literature from a communication theoretical perspective to enable the use of the method also by synthetic cells and nanoscale devices and to develop future nanonetwork applications \cite{Akyildiz2013}. To connect these networks with conventional macroscale networks, thus, to realize IoBNT, requires the implementation of seamless interfaces between the molecular and cyber domain \cite{Akyildiz2015}. This points out the need for micro/nanoscale bio-cyber gateways which can decode the molecular messages, encoded into concentration or type of molecules, process and send the decoded information to a macroscale network node through a wireless link.
\begin{figure}[!t]
\centering
\includegraphics[width=7cm]{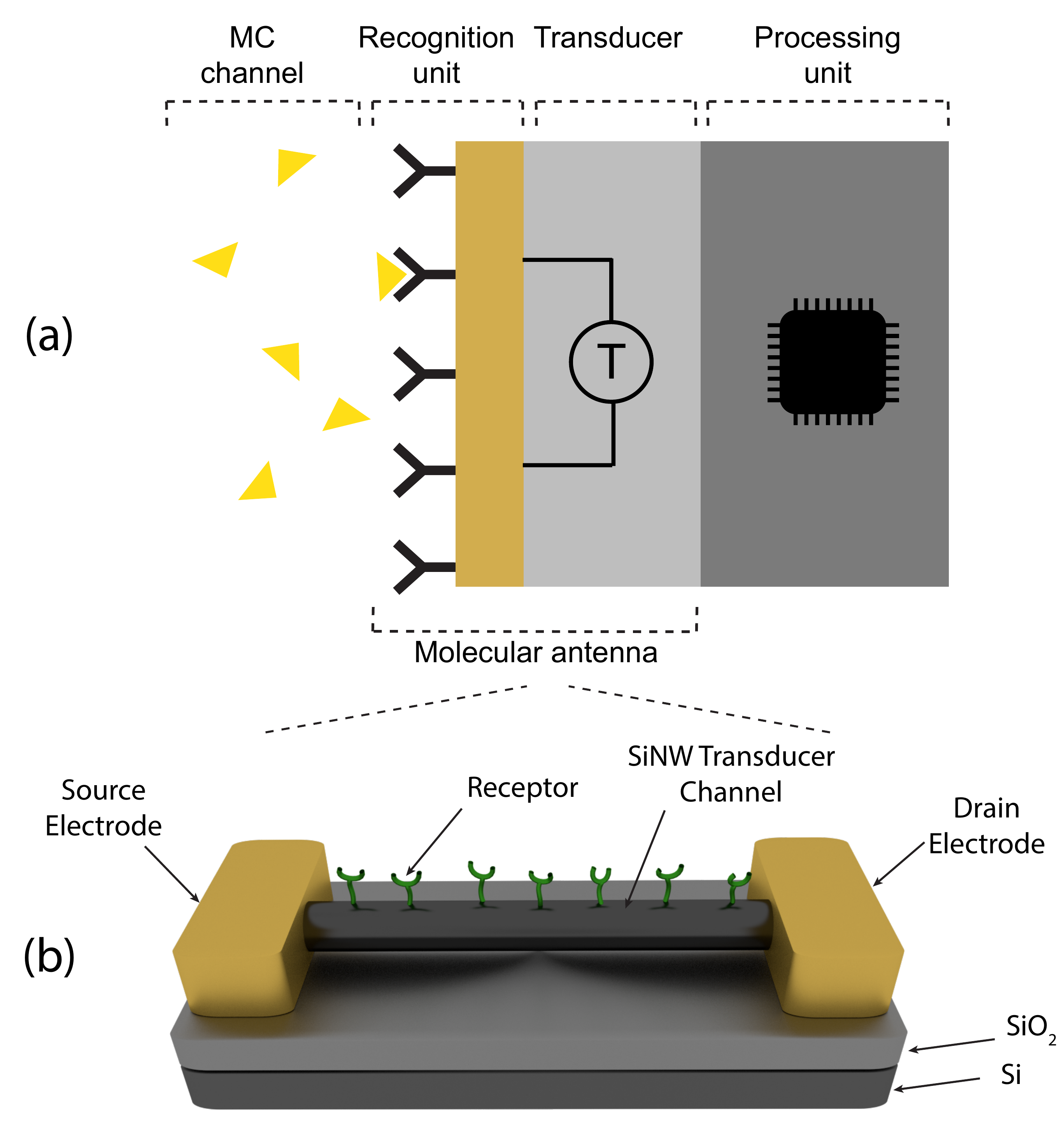}
\caption{(a) Functional units at the molecular front end of an interface, and (b) SiNW FET-based molecular antenna.}
\label{fig:MCreceiver}
\end{figure}

In \cite{Kuscu2015b}, we have comprehensively reviewed the state-of-the-art electrical biosensing approaches in the literature to determine the most promising approaches for designing a molecular communication receiver at nanoscale. Based on that study, in this paper, we take the first step towards realizing the bio-cyber interfaces and propose the use of Silicon Nanowire (SiNW) Field Effect Transistor (FET) based biosensors, i.e., SiNW bioFETs, as molecular antennas which can detect the concentration of received molecules and transduce the molecular signals into electrical form. The transduction of biochemical signals into electrical signals, with a molecular antenna conceptually depicted in Fig. \ref{fig:MCreceiver}(a), would provide the gateway with the capabilities of fast signal processing, and connecting to the cyber-networks through electromagnetic wireless communications, probably operating in the THz-band.

BioFETs are similar to the conventional FETs with the exception of an additional biorecognition layer that is capable of selectively binding the target  molecules \cite{Poghossian2014}. This layer consists of receptor molecules tethered on the surface of the FET channel, and replaces the gate electrode of conventional FETs, as shown in Fig. \ref{fig:MCreceiver}(b). Binding of ligands with intrinsic charges to the surface receptors results in accumulation or depletion of the carriers on the semiconductor channel, and modulates the channel conductance and current. Hence, the output current becomes a function of the ligand density and the amount of ligand charges. Label-free, continuous and in situ sensing of molecules by not requiring any complicated processes, such as the use of macroscale equipments for readout and processing operations, makes BioFETs a natural candidate for the molecular antenna unit of a bio-cyber interface gateway.

Several ligand-receptor pairs, e.g., antibody-antigen, aptamer-natural ligand, natural ligand/receptors, have proven suitable for the operation of bioFETs \cite{Rogers2000}. Various types of semiconductors, such as SiNW, Carbon Nanotube (CNT) and graphene, can be used as the FET channel, i.e., transducer channel \cite{Poghossian2014}. However, the literature is currently dominated by the SiNW bioFETs due to easier and controllable construction of SiNWs, leading us to focus on SiNW bioFETs \cite{Nair2007}.

Although, there is a vast number of experimental works on SiNW bioFETs and a few theoretical studies focusing on the noise sources effective on bioFETs \cite{Deen2006} \cite{Rajan2013}, the literature is lack of a comprehensive model that can enable the theoretical analysis and optimization of these devices from communication perspective. In this study, we develop a deterministic and noise models for the biorecognition and transduction operations of SiNW bioFETs when utilized as molecular antennas exposing to concentration-encoded molecular messages. The model enables us to analytically derive the SNR of the electrical signal at the antenna output, and analyze the effect of various system parameters on the antenna's performance.

The remainder of the paper is organized as follows. In Section II, we present the deterministic and noise models for the antenna. The results of performance evaluation are given in Section III. Finally, Section IV concludes the paper.

\section{SiNW FET-Based Molecular Antenna Model}
\label{Numerical}
In this section, we develop a deterministic model for the signal flow from the capture of molecules by the biorecognition layer to the output current in the transducer channel. We also provide models for the main noise sources to derive the SNR of the antenna's output current. We assume that the molecular messages are represented by different levels of ligand concentration.

\subsection{Deterministic Model}
We begin with the dynamics between the information-carrying ligand molecules and the receptor molecules tethered on the surface of the NW transducer channel. In order to derive analytical expressions, we make the following assumptions:
\begin{itemize}
\item Diffusion of ligands are assumed to be fast enough such that the reception is not mass transport limited. This implies that the ligands are homogenously distributed in the reception space, i.e., in the vicinity of surface receptors, and each of the receptors is exposed to the same concentration of ligands. 
\item Ligand concentration is assumed to be much higher than the surface receptor concentration such that the ligand concentration in the reception space remains almost constant despite the ligand-receptor reactions taking place.
\end{itemize}
These assumptions are prevalent in molecular communications studies \cite{Akyildiz2013}, and lead to a pseudo-first order ligand-receptor dynamics, where the first time derivative of the number of bound receptors $dN_B(t)/dt$ can be expressed through the following differential equation:
\begin{equation}
\frac{d N_B (t)}{dt} = k_+ c_L^R(t)(N_R - N_B(t)) - k_- N_B(t), \label{eq:dNb/dt}
\end{equation}
where $k_+$ and $k_-$ are the intrinsic association and dissociation rate constants of the receptor-ligand complex, respectively. $N_R$ is the total number of receptors on the bioFET surface, and $c_L^R(t)$ is the ligand concentration in the reception space.
\begin{figure}[!t]
\centering
\includegraphics[width=6cm]{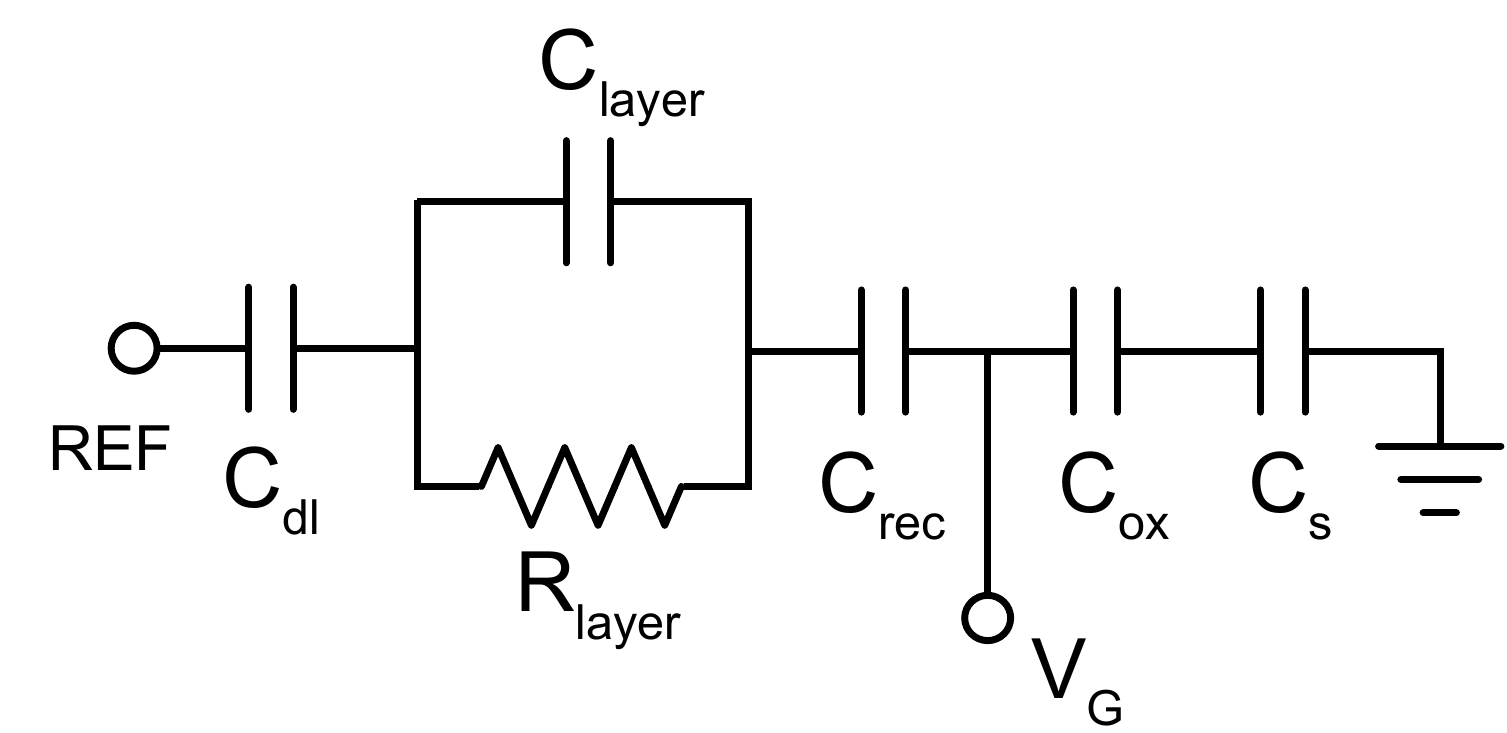}
\caption{Equivalent circuit model for the transducer of the SiNW FET-based MC receiver \cite{Deen2006} \cite{Spathis2015}. \emph{REF} denotes the reference electrode.}
\label{fig:Circuit}
\end{figure}

Since the mass transport effects on the reaction dynamics are neglected, we can assume that a concentration change in the bulk due to successively transmitted messages from the information source is immediately reflected to the concentration in the reception space. Therefore, we can neglect the transient phase between different levels of ligand concentration such that $c_L^R(t) = c_i$ for $t \in [t_i, t_i+1/B)]$, where $c_i$ is the ligand concentration level in the reception space corresponding to the $i$th message, $t_i$ is the transition time from the $(i-1)$th message to the $i$th message in the reception space, and $1/B$ is the symbol duration with $B$ being the symbol transmission rate. In other words, the biorecognition layer is assumed to be exposed to a constant concentration $c_i$ representing the $i$th symbol for $t \in [t_i, t_i+1/B)]$.

Given the initial condition $N_B(t_i-\epsilon) = N_{B,i-1}$ with $\epsilon \rightarrow 0$, the solution of the differential equation \eqref{eq:dNb/dt} can be given as \cite{Berezhkovskii2013}
\begin{multline}
N_B(t) = N_{B,i}^{ss} + \left( N_{B,i-1} - N_{B,i}^{ss} \right) e^{-(k_+ c_i + k_-) (t-t_i)}  \\ \text{for}\ t \in [t_i, t_i+1/B),
\end{multline}
where $N_{B,i}^{ss}$ is the number of receptors at steady-state, i.e., when $dN_B(t)/dt = 0$. We infer from this equation that although the ligand concentration level immediately changes in the reception space, it takes a certain time for the receptors to adapt the concentration level of the new message. The time to reach steady-state is governed by the reaction timescale $\tau_B = (k_+ c_i + k_-)^{-1}$; thus, for higher ligand concentrations, the adaptation time of the recognition layer decreases. The number of occupied receptors at steady-state is given by
\begin{equation}
N_{B,i}^{ss} = \frac{k_+ c_i}{k_+ c_i + k_-} N_R = \frac{c_i}{c_i + K_D} N_R. \label{ss}
\end{equation}

We assume that the gateway samples the receptor states at steady-state; thus, the number of occupied receptors corresponding to the $i$th message can be given as $N_{B,i} = N_{B,i}^{ss}$.

The charged ligands bound to the surface receptors induce opposite charges on the gate of bioFET. The mean amount of charge generated for the $i$th message is given by $Q_i = N_{B,i} \, N_e \, q_{eff}$, where $N_e$ is the number of free electrons per ligand molecule. $q_{eff}$ is the mean effective charge that can be reflected to the gate by a single electron of a ligand molecule in the presence of Debye screening. The mean effective charge of a free ligand electron as observed by the transducer is degraded as the distance between the ligand electron and the transducer increases. The relation is given by $q_{eff} = q \times exp(-r/\lambda_D)$, where $q$ is the elementary charge, and $r$ is the average distance of ligand electrons in the bound state to the transducer's surface \cite{Rajan2013}, which is assumed to be equal to the average length of receptor molecules, i.e., $r = L_R$. $\lambda_D$ is the Debye length which quantizes the ionic strength of the solution according to the following relation
\begin{equation}
\lambda_D = \sqrt{\frac{\epsilon_R k_B T}{2 N_A q^2 c_{ion}}}, \label{eq:debye}
\end{equation}
where $\epsilon_R$ is the dielectric permittivity of the fluidic medium, $k_B$ is the Boltzmann's constant, $T$ is the temperature, and $N_A$ is Avogadro's number, $c_{ion}$ is the ionic concentration of the medium \cite{Rajan2013}.

The induced charges on the gate are translated into the gate voltage through the equivalent circuit of the transducer \cite{Deen2006} \cite{Spathis2015}, which is demonstrated in Fig. \ref{fig:Circuit}. By neglecting the current through $R_{layer}$, i.e., resistance of the layer of bound ligands, which is on the order of tens of $G \Omega$s \cite{Spathis2015} \cite{Bang2008}, the gate voltage resultant from the bound ligands can be written as $V_{G,i} = Q_i/C_{eq,i}$, where the overall capacitance of the equivalent circuit $C_{eq,i}$ is expressed by
\begin{multline}
C_{eq} = \left((C_{ox} W L)^{-1} + (C_{s} W L)^{-1}\right)^{-1}\\ + \left(C_{rec}^{-1} + C_{layer,i}^{-1} + (C_{dl} W L)^{-1}\right)^{-1}.
\end{multline}
Here, $C_{ox}$, $C_{s}$ and $C_{dl}$ are the oxide, the semiconductor, i.e. SiNW, and the double layer capacitances per unit area, respectively; $C_{rec}$ and $C_{layer,i}$ are the capacitances of the receptor layer and the layer of bound ligands when the $i$th message is received, respectively; and $W$ and $L$ are the width and length of the transducer's active region. $C_{ox} = \epsilon_{ox}/t_{ox}$ with $\epsilon_{ox}$ and $t_{ox}$ being the permittivity and the thickness of the oxide layer. $C_{rec} = N_R \times C_{mol,R}$, $C_{layer,i} = N_{B,i} \times C_{mol,L}$ with $C_{mol,R}$ and $C_{mol,L}$ being the capacitance of a single receptor and a single ligand molecule, respectively.

\begin{figure}[!t]
\centering
\includegraphics[width=6cm]{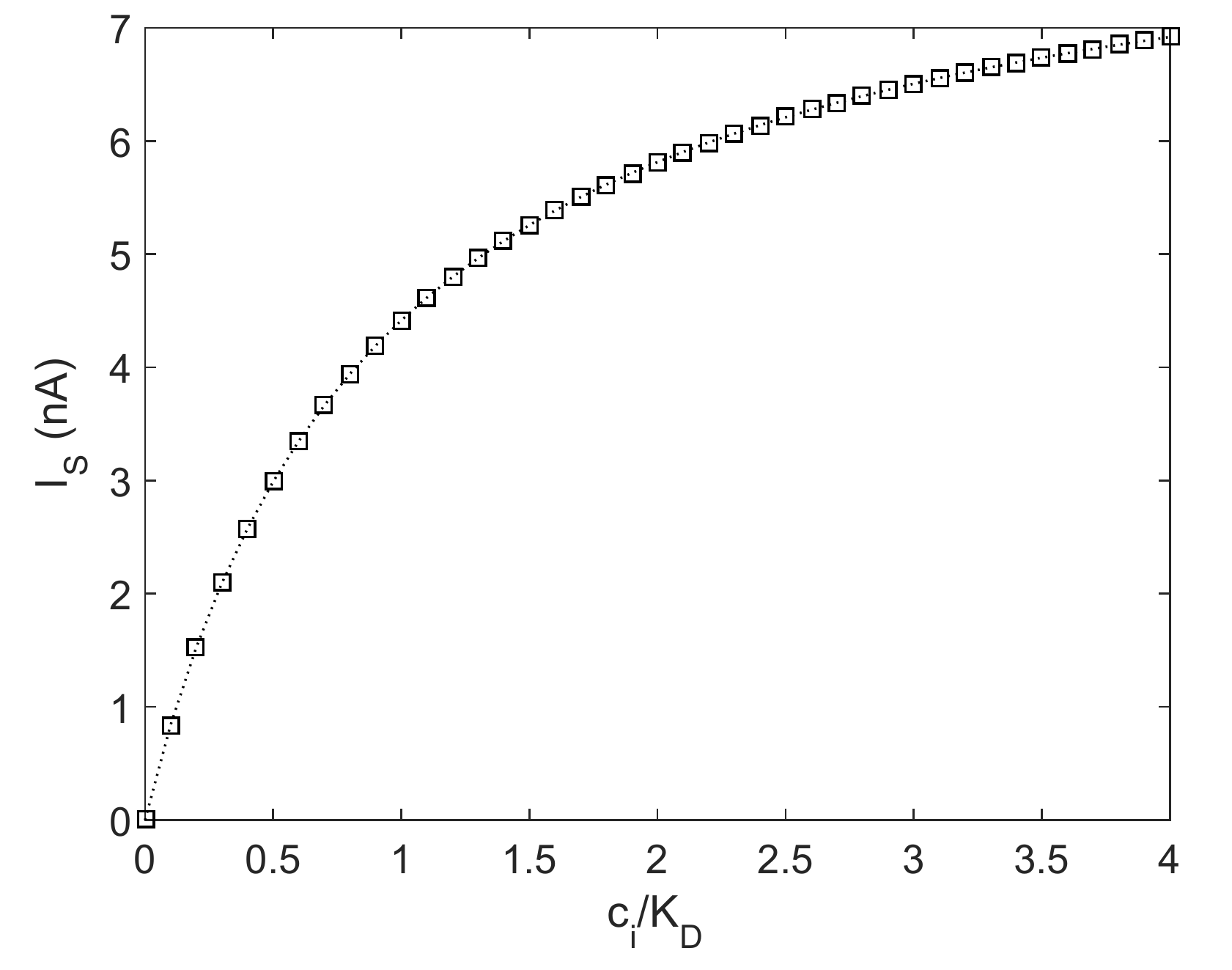}
\caption{Mean output current $I_S$ of molecular antenna for varying $c_i$, as simulated using the parameters in Table \ref{table:parameters}.}
\label{fig:response}
\end{figure}

The induced gate voltage is reflected into a variation in the current flowing through the transducer channel $I_{M,i} = I_{S,i} + I_0$, where $I_0$ is the bias current which is assumed to be constant and independent of the gate voltage $V_{G,i}$ \cite{Rajan2013-2}; and $I_{S,i}$ is the current resultant from the bound ligands during the reception of $i$th message. We are interested in $I_{S,i}$ since it is modulated by ligand concentration: $I_{S,i} = V_{G,i}\, g_m$, where $g_m$ is NW channel transconductance, expressed by
\begin{equation}
g_m = (W/L) \mu_{eff} C_{ox} V_{DS}, \label{eq:conductance}
\end{equation}
where $\mu_{eff}$ is the effective carrier mobility in the transducer channel, and $V_{DS}$ is the drain-to-source voltage, which is assumed to be held constant \cite{Rajan2013-2}. The processor unit in the receiver uses the signal $I_{S,i}$, which is plotted in Fig. \ref{fig:response} for the parameters in Table \ref{table:parameters}, to infer the incoming ligand concentration $c_i$; and thus the molecular message $i$, based on a predefined CSK scheme.

\subsection{Noise Model}
We investigate the contributions of three main noise sources, which shows zero-mean additive behavior on the resultant current signal:
\begin{inparaenum}[\itshape i\upshape)]
\item receptor noise which is originating from the stochastic ligand-receptor bindings on the biorecognition layer;
\item thermal noise resulting from the stochastic movement of free electrons on the bound ligands; and
\item $1/f$ noise of the transducer, which is intrinsic to all electronic devices, and caused by the defects and traps on the transducer channel.
\end{inparaenum}

\subsubsection{Receptor Noise}
The number of occupied receptors fluctuates at any time, even at the steady-state, due to the stochastic receptor-ligand binding dynamics. To derive the resultant current noise associated with the binding fluctuations, we need to express the dynamics with a stochastic model. The pseudo-first order stochastic dynamics is well described by the following Markov Chain (MC):
\begin{equation}
\ce{0 <=>[\alpha_0][\beta_1] 1 <=>[\alpha_1][\beta_2] 2\, \, \, \ldots \, \, \,N_R - 1 <=>[\alpha_{N_R-1}][\beta_N] N_R},
\end{equation}
where the states of the MC are denoting the number of occupied receptors. The state-dependent transition rates are given as $\alpha_n = (N_R - n) k_+ c_i$ and $\beta_n = n k_-$. The corresponding forward equations can be written as \cite{Berezhkovskii2013}
\small
\begin{equation}
  \begin{aligned}
  \frac{dP_0(t)}{dt} &= -\alpha_0 P_0(t) + \beta_1 P_1(t),  \\
  \frac{dP_n(t)}{dt} &= -\alpha_{n-1} P_{n-1}(t) - (\alpha_n + \beta_n) P_n(t) + \beta_{n+1} P_{n+1}(t), \\
  \frac{dP_{N_R}(t)}{dt} &= -\alpha_{N_R-1} P_{N_R-1}(t) + \beta_N P_N(t),
  \end{aligned}
\end{equation}
\normalsize
where $P_n(t)$ is the probability of being in state $n$ at time $t$. Since the receiver samples the receptor states at steady-state, we are interested in the steady-state distribution $P_n^{ss}$, which can be calculated by setting all time derivatives to zero:
\begin{equation}
P_n^{ss} = \frac{N_R!}{n!(N_R-n)! } \frac{(k_+ c_i)^n k_-^{N_R - n}}{(k_+ c_i + k_-)^{N_R}}.
\end{equation}
The mean and variance of the number of occupied receptors at steady-state are then given by \cite{Berezhkovskii2013}
\begin{equation}
\overline{N_{B,i}} = \sum_{N_{B,i}=0}^{N_R} N_{B,i} P_{N_{B,i}}^{ss} = \frac{k_+ c_i}{k_+ c + k_-} N_R, \label{eq:mean}
\end{equation}
\begin{equation}
Var(N_{B,i}) = \overline{N_{B,i}^2} - (\overline{N_{B,i}})^2 = \frac{k_- k_+ c_i}{(k_+ c_i + k_-)^2} N_R. \label{eq:variance}
\end{equation}
Equation \eqref{eq:mean} is exactly the same as the steady-state solution \eqref{ss} of the deterministic differential equation \eqref{eq:dNb/dt}. The autocorrelation function for the stationary fluctuations at the steady-state can be approximated with a single exponential \cite{Bialek2005}:
\begin{equation}
R(\tau) = Var(N_{B,i}) e^{-\frac{\tau}{\tau_B}}, \label{eq:acf}
\end{equation}
where the characteristic timescale of ligand-receptor binding $\tau_B = (k_+ c_i + k_-)^{-1}$ is also being the correlation time of binding noise. The Fourier Transform of \eqref{eq:acf} gives the Power Spectral Density (PSD) of the fluctuations:
\begin{equation}
S_{\Delta N_B}(f) = Var(N_{B,i}) \frac{2 \tau_B}{1+(2 \pi f \tau_B)^2}. \label{eq:SNb}
\end{equation}

Given the noise PSD for number of bound receptors, the PSD of fluctuations in voltage $V_g$ can be written as
\begin{equation}
S_{\Delta V_G^B}(f) = S_{\Delta N_B}(f) V_m^2, \label{eq:VGB}
\end{equation}
where $V_m = (N_e q_{eff})/C_{eq,i}$ is the mean deviation in the gate voltage resulting from binding of a single ligand. As can be inferred from Equations \eqref{eq:SNb} and \eqref{eq:VGB}, the gate voltage noise PSD associated with the binding fluctuations is Lorentzian with a critical frequency of $f_B = \tau_B^{-1}$.

\subsubsection{Thermal Noise}
Random diffusion of free electrons results in thermal noise on the resistive layer of bound ligands. Since the extent of the field effect strongly depends on the distance of the stimulating ligand electrons to the NW surface due to the Debye electrolyte screening, the uncertainty in the location of electrons are reflected into fluctuations in the gate voltage of bioFET. Using the thermal noise model derived in \cite{Spathis2015}, the PSD of voltage fluctuations on the layer of bound ligands can be expressed by $S_{\Delta V_{R_{layer,i}}^T} = 4 k T R_{layer,i}$, where $R_{layer,i}$ is the resistance of the layer of bound ligands when the $i$th message is received, $k$ is the Boltzmann constant, $T$ is the temperature. The fluctuations in the voltage across $R_{layer,i}$ is reflected into the gate voltage through the RC network shown in Fig. \ref{fig:Circuit}. Using the transfer function of the RC network, the PSD of the resultant thermal noise contribution on the gate voltage can be written as \cite{Spathis2015}
\begin{equation}
S_{\Delta V_G^T}(f) = \frac{S_{\Delta V_{R_{layer,i}}^T}}{ 1 + \left(2 \pi R_{layer,i} (C_{layer,i}+C_{eq}') f\right)^2 },
\end{equation}
where
\small
\begin{equation}
C_{eq}' = \left((C_{dl}WL)^{-1} + C_{rec}^{-1} + (C_{ox}WL)^{-1} + (C_sWL)^{-1}\right)^{-1}.
\end{equation}
\normalsize

\begin{figure}[!t]
\centering
\includegraphics[width=7cm]{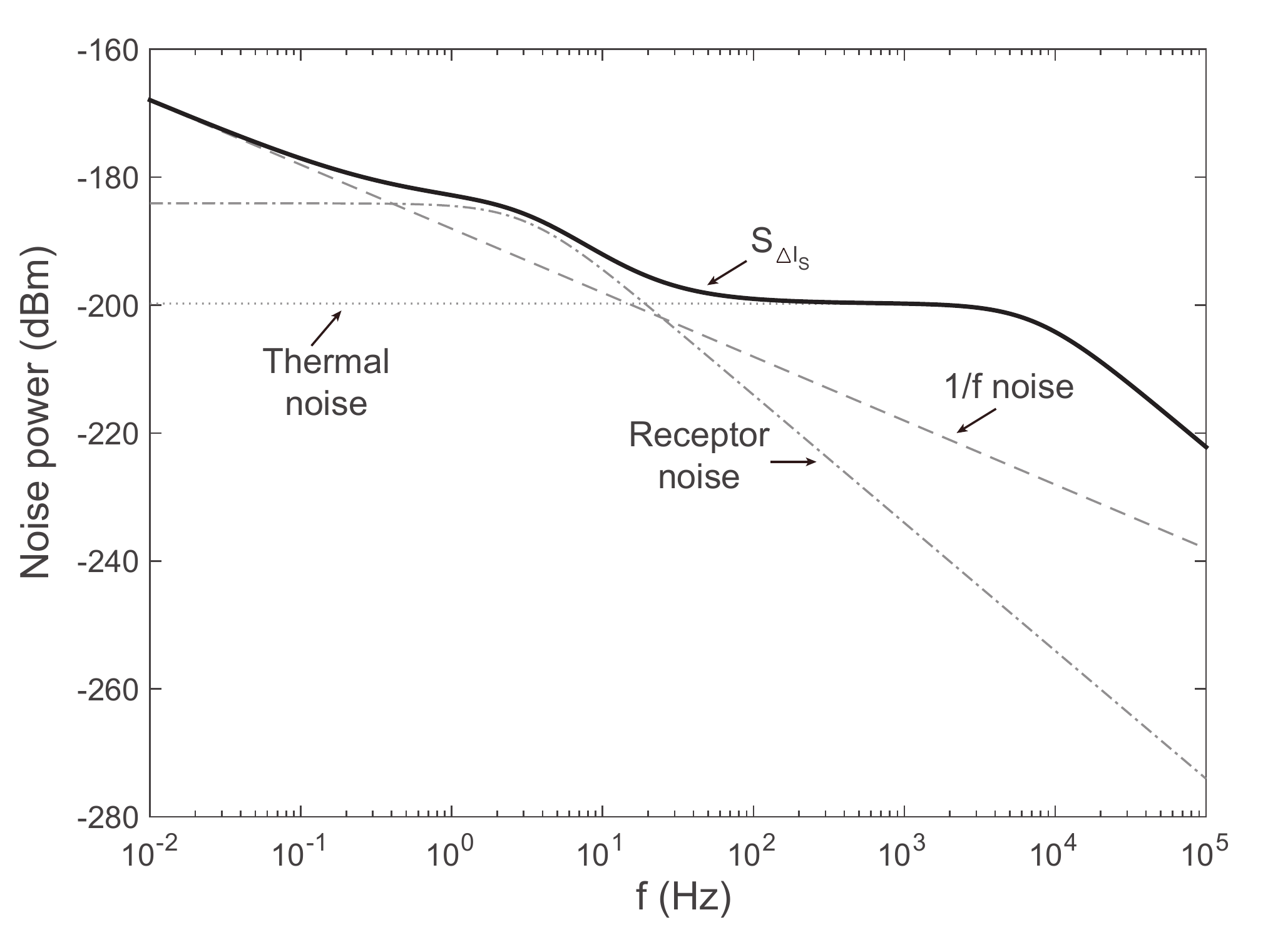}
\caption{Noise PSD for SiNW FET-based molecular antenna including the contributions of different types of noise sources.}
\label{fig:PSDtotal}
\end{figure}

The thermal noise associated with the ligand electrons, which is the PSD of white noise colored by the RC network transfer function, has also a Lorentzian PSD with critical frequency $f_T = \left(R_{layer,i} (C_{layer,i} + C_{eq'})\right)^{-1}$.

\subsubsection{Flicker Noise}

As in all transistor devices, low-frequency operation of bioFET-based molecular antenna is suffered from $1/f$ noise. Although the origin of flicker noise and its full analytical model are still open issues, there are several models, including the well-known Hooge's model, that approximate the noise power in frequency domain \cite{Rajan2013-2}.

In this paper, we use the number fluctuation model, which provides a more accurate approximation compared to Hooge's model, attributing the source of $1/f$ noise to the random generation and recombination of charge carriers due to the defects and traps in the transducer channel \cite{Rajan2013-2}. Fluctuations due to random generation and recombination of individual charge carriers, which follow Lorentzian spectrum, sum up to construct $1/f$ noise. The model expresses the resultant gate voltage-referred noise PSD as follows,
\begin{equation}
S_{\Delta V_G^F}(f) = \frac{\lambda k T q^2 N_t}{W L C_{ox}^2 f},
\end{equation}
where $\lambda$ is the characteristic tunneling distance, and $N_t$ is the trap density of the NW channel. $1/f$ noise is independent of the received signals, and shows an additive behavior on the overall gate voltage fluctuations \cite{Spathis2015}. Theoretically, $1/f$ noise does not have a low frequency cutoff, and has infinite power at zero frequency. However, in experimental studies with a finite measurement time, a finite variance for $1/f$ noise is observed. The reason is related to the low frequency cutoff set by the observation time $T_{obs}$ \cite{Niemann2013}. Considering that the received molecular signals are at the baseband, to be able to calculate the total noise power, we assume one-year operation time for the antenna such that the low cutoff frequency is $f_L = 1/T_{obs} \approx  3 \times 10^{-8}$Hz. At frequencies lower than $f_L$, the noise is assumed to show a white noise behavior, i.e., $S_{\Delta V_G^F}(f) = S_{\Delta V_G^F}(f_L)$ for $f < f_L$.

\subsubsection{Total Noise Power and SNR}
The receptor noise resultant from slow receptor-ligand dynamics and the thermal noise resultant from the fast electron diffusion can be assumed uncorrelated with each other, because they are largely separated in the frequency domain \cite{Spathis2015}. Therefore, including the additive $1/f$ noise, the overall PSD of the gate voltage referred noise can be expressed as
\begin{equation}
S_{\Delta V_G}(f) = S_{\Delta V_G^B}(f) + S_{\Delta V_G^T}(f) + S_{\Delta V_G^F}(f).
\end{equation}
Gate voltage fluctuations are reflected into channel current noise by $S_{\Delta I_S}(f) = S_{\Delta V_G}(f) g_m^2$, where the transconductance $g_m^2$ is given in \eqref{eq:conductance}. Assuming a resistance of 1$\Omega$ for the channel, SNR for the antenna output can be computed by
\begin{equation}
SNR = \frac{I_S^2}{\int_{-\infty}^{\infty} S_{\Delta I_S}(f) df}. \label{eq:SNRI}
\end{equation}

\begin{figure}[!t]
\centering
\includegraphics[width=6.75cm]{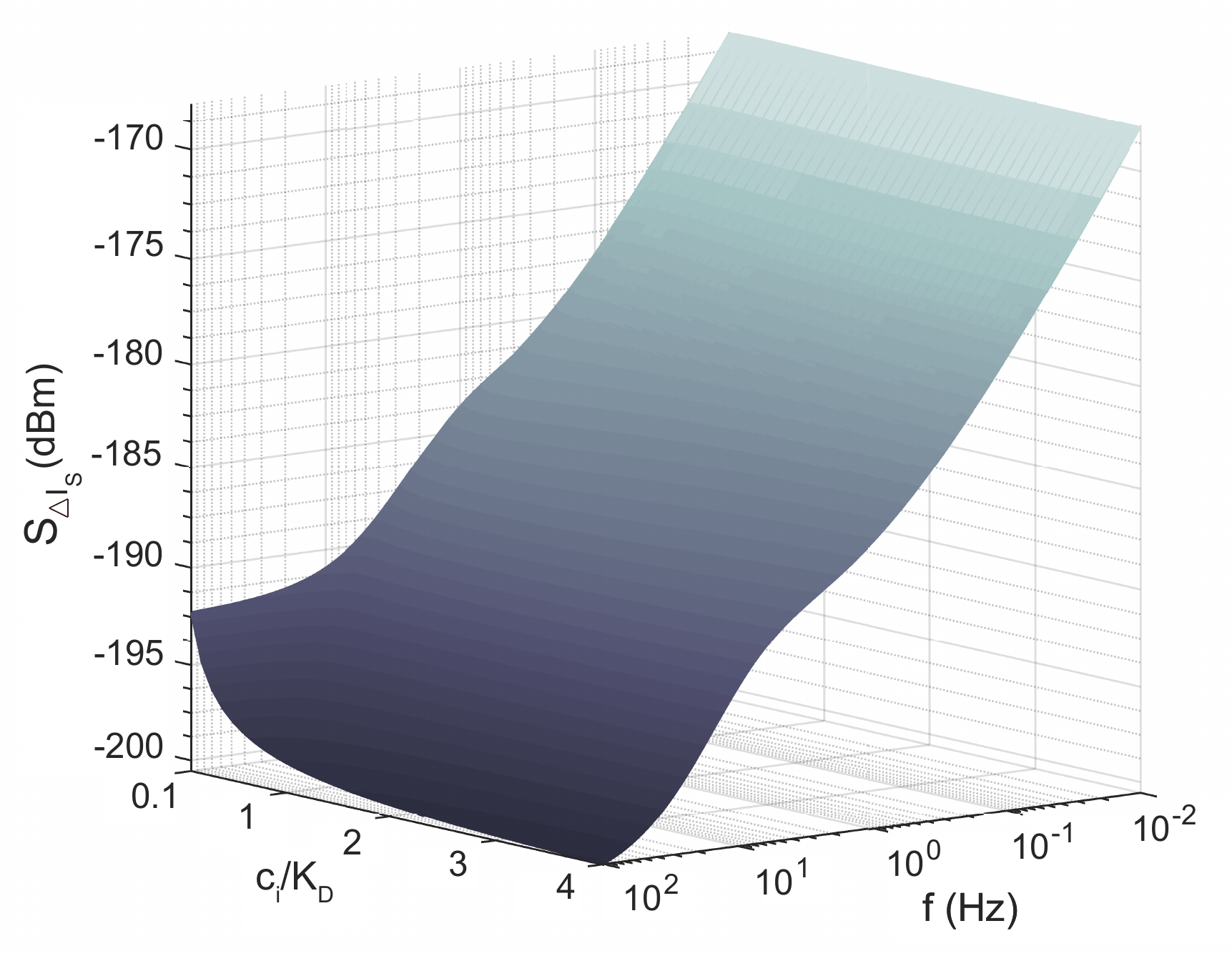}
\caption{Noise PSD with varying ligand concentration $c_i$.}
\label{fig:PSDci}
\end{figure}

\section{Performance Evaluation}
In this section, we provide the results of the simulations for the deterministic and noise model under different settings. The default values for the main system parameters used in the simulations are listed in Table \ref{table:parameters}, where the sources of the data are also noted. The parameter values are selected assuming that the MC is exposed to the physiological conditions, and the receptor-ligand pairs correspond to antigen-antibody, aptamer-DNA, or aptamer-protein pairs.

\begin{figure*}[!t]
 \centering
   \subfigure[]{
 \includegraphics[width=4.2cm]{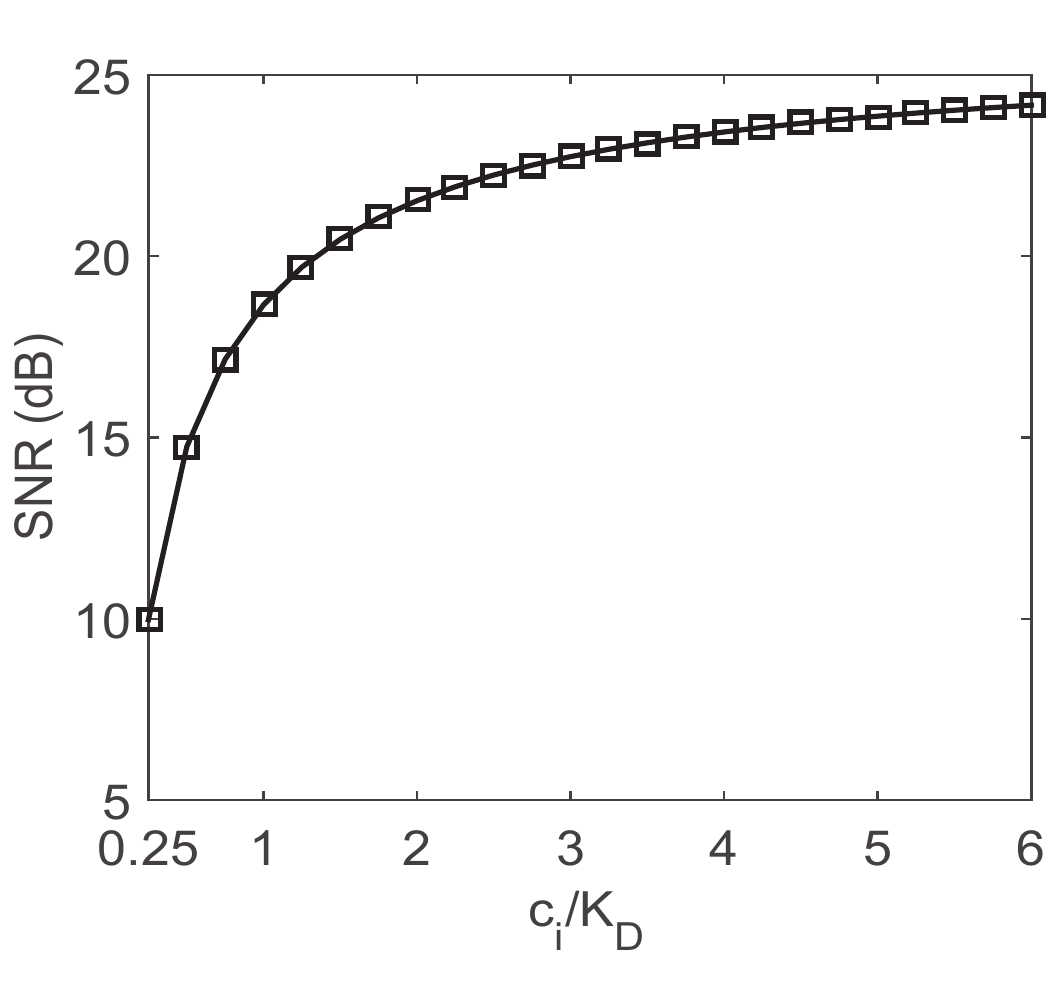}
 \label{fig:SNRci}
 }
  \subfigure[]{
 \includegraphics[width=4.2cm]{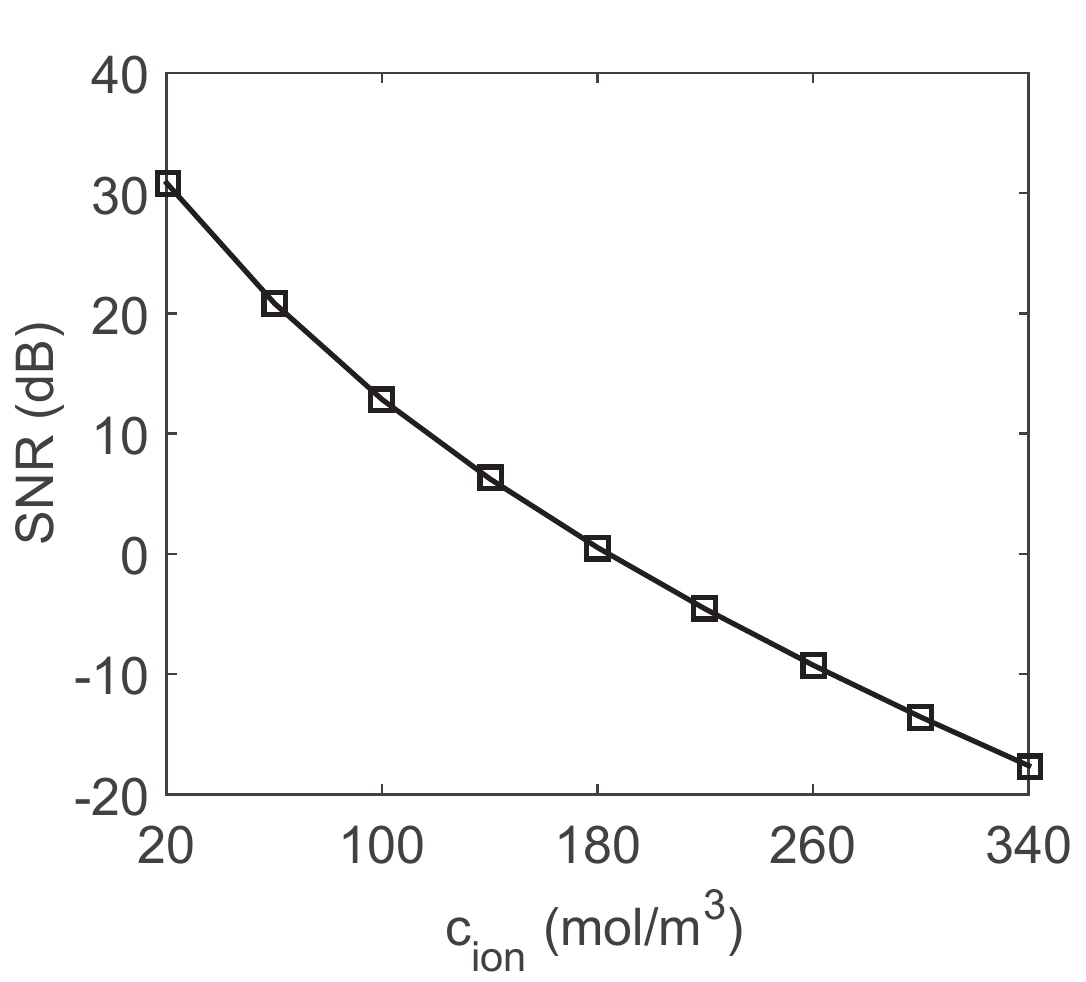}
 \label{fig:SNRcion}
 }
    \subfigure[]{
 \includegraphics[width=4.2cm]{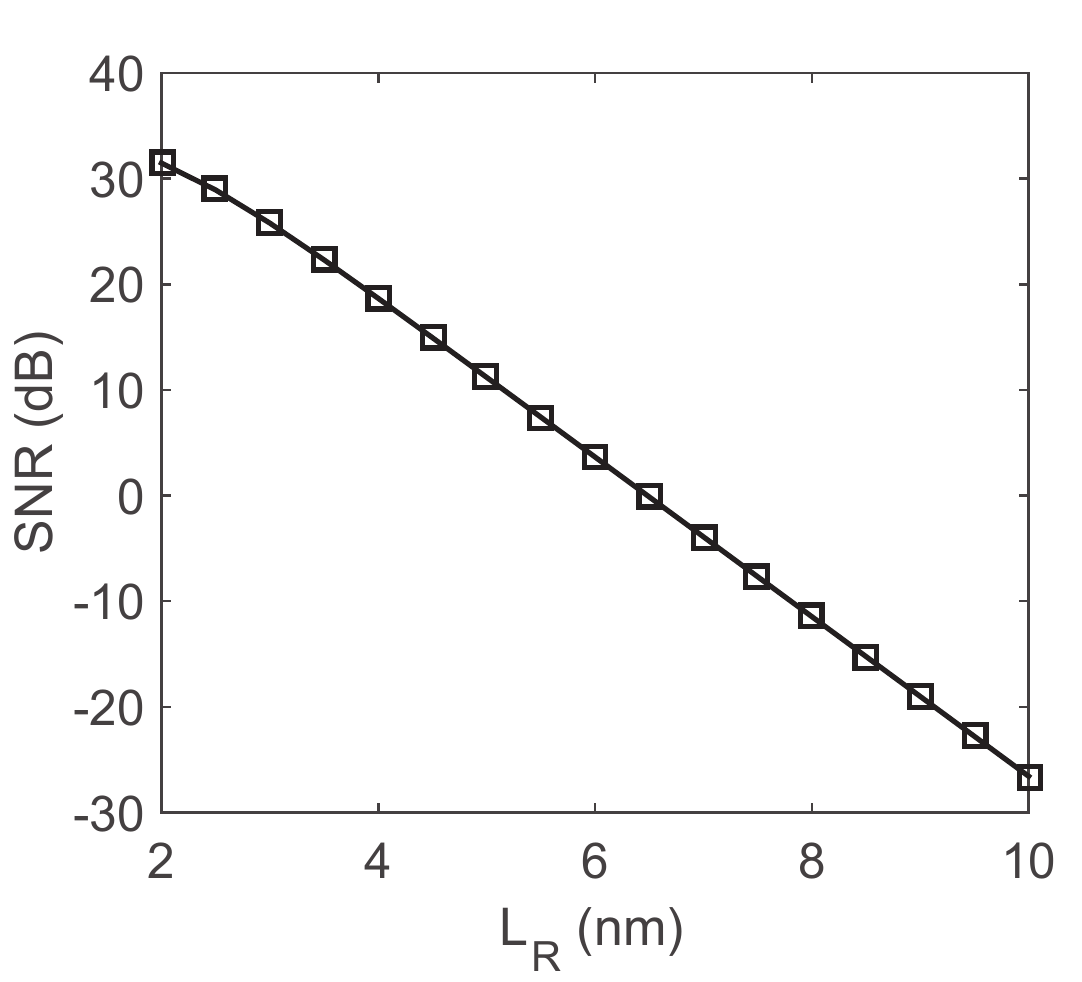}
 \label{fig:SNRLr}
 }
  \subfigure[]{
 \includegraphics[width=4.2cm]{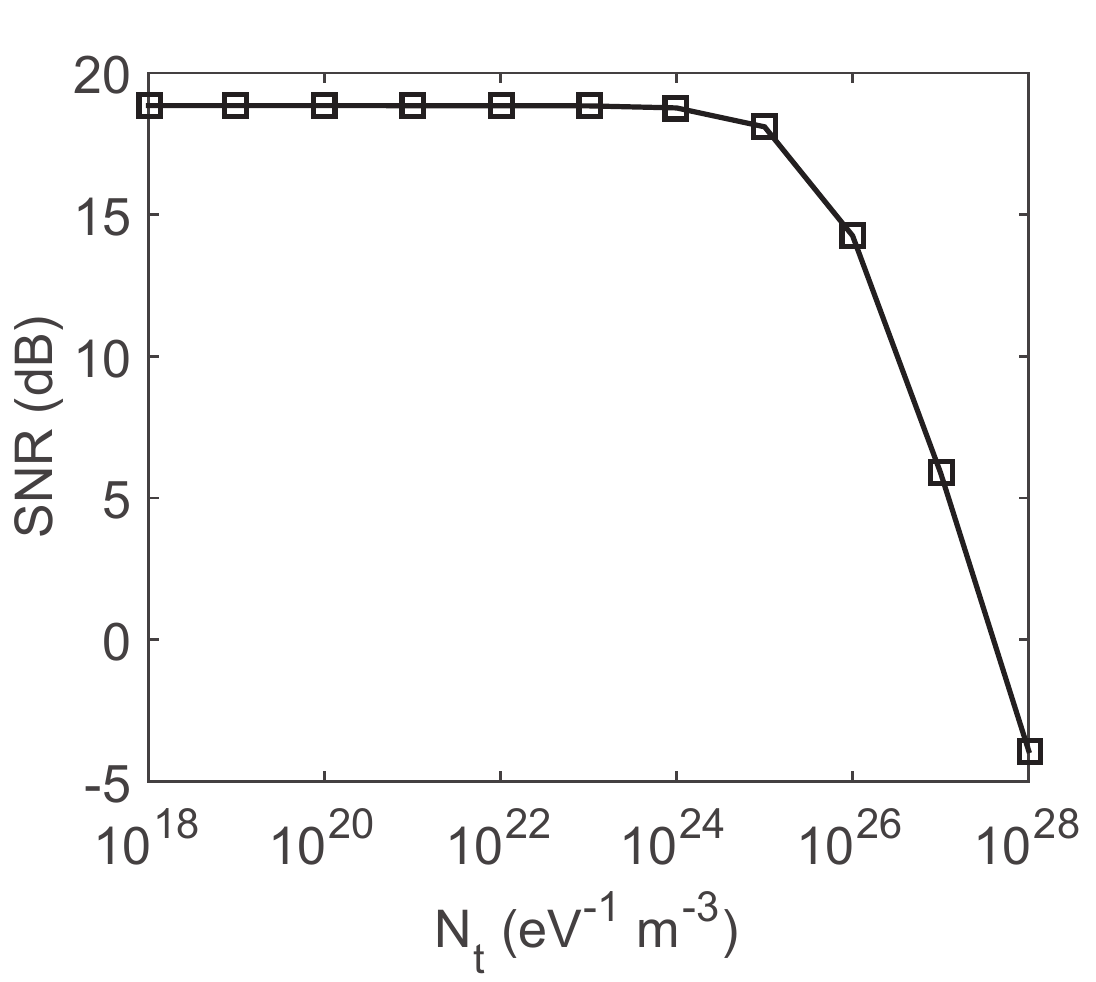}
 \label{fig:SNRNt}
 }
 \caption{SNR with (a) varying concentration $c_i$ of information-carrying ligands, (b) varying ionic strength $c_{ion}$, (c) varying receptor length $L_R$, and (d) varying trap density $N_t$ in the transducer channel.}
 \label{fig:SNRcicion}
 \end{figure*}

We first evaluate the PSD of resultant current noise given in Fig. \ref{fig:PSDtotal}. As is seen, the frequency domain is virtually divided into three regions in each of which one of the three noise sources is prevailing. At low frequencies $f \ll f_B \sim 2$Hz, $1/f$ noise is dominating over the receptor and thermal noise. As the frequency gets higher, the receptor noise becomes dominant around $f_B$. The critical frequency of the thermal noise $f_T$ will be on the order of $10$kHz for the settings given in Table \ref{table:parameters}. At frequencies higher than $10$Hz, the power of receptor and $1/f$ noise substantially attenuates, and the noise is dominated by the contribution of the thermal noise.
\begin{table}[!b]\scriptsize
\centering
\caption{Simulation Parameters}
\begin{tabular}{ l | l }
   \hline \hline
   Width and length of active region ($W \times L$) & $0.1 \times 5$ ($\mu m \times \mu m$) \cite{Rajan2011a} \\ \hline
   Tunneling distance ($\lambda$) & $0.05$ ($nm$) \cite{Deen2006} \\ \hline
   Temperature ($T$) & $300$ ($K$) \\ \hline
   Relative permittivity of oxide layer ($\epsilon_{ox}/\epsilon_0$) & $3.9$ \cite{Deen2006} \\ \hline
   Thickness of oxide layer ($t_{ox}$) & $17.5$ ($nm$) \cite{Deen2006} \\ \hline
   Trap density ($N_t$) & $2.3 \times 10^{24}$ ($eV^{-1} m^{-3}$) \cite{Rajan2011a} \\ \hline
   Effective mobility ($\mu_{eff}$) & $16 \times 10^{-3}$ ($m^2 V^{-1} s^{-1}$) \cite{Deen2006} \\ \hline
   Drain-source voltage ($V_{DS}$) & $0.1$ ($V$) \cite{Rajan2013-2} \\ \hline
   Relative permittivity of medium ($\epsilon_R/\epsilon_0$) & $78$ \cite{Hediger2012} \\ \hline
   Ionic strength of medium ($c_{ion}$) & $70$ ($mol/m^3$) \cite{Okada1990} \\ \hline 
   Average number of electrons in a ligand ($N_e$) & $3$ \cite{Fritz2002} \\ \hline  
   Length of receptor ($L_R$) & $4$ ($nm$) \cite{Song2008} \\ \hline   
   Binding rate ($k_+$) & $0.2 \times 10^{-18}$ ($m^3 s^{-1}$) \cite{Akyildiz2013} \\ \hline
   Unbinding rate ($k_-$) & $10$ ($s^{-1}$) \cite{Akyildiz2013} \\ \hline
   Ligand concentration in reception space ($c_i$) & $K_D$  \\ \hline
   Concentration of receptors on the surface ($c_R$) & $10^{16}$ ($m^{-2}$) \cite{Rajan2013-2} \\ \hline
   Molecular resistance ($R_{mol,L}, R_{mol,R}$) & $4 \times 10^{14}$ ($\Omega$) \cite{Bang2008} \\ \hline
   Molecular capacitance ($C_{mol,L}, C_{mol,R}$) & $2 \times 10^{-20}$ ($F$) \cite{Shorokhov2011} \\ \hline
   Capacitance of dielectric layer ($C_{dl}$) & $5 \times 10^{-2}$ ($F/m^2$) \cite{Landheer2005} \\ \hline
   Capacitance of silicon ($C_s$) & $2 \times 10^{-3}$ ($F/m^2$) \cite{Landheer2005} \\ \hline
\end{tabular}
\label{table:parameters}
\end{table}

As the bandwidth of the received signal is expected to be at most on the order of Hz \cite{Akyildiz2013}, it would be sufficient for the gateway to sample the receptor states at not more than $10$Hz. In this frequency range, we can expect that only $1/f$ and receptor noise would be effective on the antenna performance.

The noise PSD is also analyzed for varying ligand concentration (normalized to dissociation constant $K_D$) corresponding to different symbols in the CSK scheme. As can be seen from Fig. \ref{fig:PSDci}, the contribution of $1/f$ noise dominating in mHz region does not vary remarkably as the input concentration changes; however, the contribution of receptor noise becomes more prevailing in lower concentrations. This is originating from the fact that the variance of number of occupied receptors at steady-state increases for lower ligand concentrations as can be inferred from \eqref{eq:variance}. The figure also clearly demonstrates that the critical frequency of the receptor noise $f_B$ increases for higher concentrations as expected. Negligible contribution of thermal noise is not evident in this frequency range.

Next, we investigate the effect of different parameters on the SNR of the antenna output, which is defined in \eqref{eq:SNRI}. SNR for varying ligand concentration corresponding to different symbols is plotted in Fig. \ref{fig:SNRci}, which clearly shows that SNR is significantly improved with increasing concentration. However, it begins to saturate at around $25$dB due to the saturation of the surface receptors for high ligand concentrations.

The effect of ionic strength of the fluidic medium on the output SNR is given in Fig. \ref{fig:SNRcion}. When the ionic concentration increases above 100mol/m$^3$, the Debye length decreases below 1nm resulting in substantial screening of ligand charge. Therefore, SNR significantly decreases with increasing ionic strength. Physiological conditions imply ionic concentrations higher than 100mol/m$^3$. To compensate the attenuation of SNR, receptors with lengths comparable to Debye length should be selected. We also investigate the effect of receptor length on the SNR when the ionic strength is 70mol/m$^3$ which makes the Debye length equal to 1.15nm. As seen in Fig. \ref{fig:SNRLr}, SNR in dB decreases linearly as the receptor length increases.

Lastly, we analyze the SNR for varying trap density which is inversely proportional to the purity of the transducer channel. Trap density increases the $1/f$ noise, which is very effective in the frequency range of the antenna's operation. As is shown in Fig. \ref{fig:SNRNt}, the effect of trap density on the $1/f$ noise, and thus, on the SNR, is evident especially for $N_t > 10^{24}$eV$^{-1}$m$^{-3}$. Fortunately, experimentally reported trap densities for SiNW bioFETs are on the order of $10^{22}$eV$^{-1}$m$^{-3}$ \cite{Rajan2011a}.

\section{Conclusion}
In this paper, we proposed the use of SiNW bioFETs as molecular antennas to realize bio-cyber interfaces between molecular and macroscale networks for the ultimate aim of enabling the IoNBT. We developed deterministic and noise models for SiNW FET-based molecular antennas to provide a theoretical optimization framework. The results of the performance evaluation revealed high SNR values at the antenna output for common system settings, and justified our proposal of utilizing SiNW bioFETs as molecular antennas.

\section*{Acknowledgment}
This work was supported in part by the European Research Council (ERC) under grant ERC-2013-CoG $\#616922$.

\end{document}